\documentstyle[preprint,aps]{revtex}
\input epsf
\begin{document}

\draft

\preprint{draft mmm00 8/12/96}

\title{Quantum Computation and Spin Physics}

\author{David P. DiVincenzo}
\address{
IBM Research Division\\
Thomas J. Watson Research Center\\
P. O. Box 218\\
Yorktown Heights, NY 10598 USA\\
}

\date{\today}

\maketitle

\begin{abstract}
A brief review is given of the physical implementation of quantum
computation within spin systems or other two-state quantum systems.
The importance of the controlled-NOT or quantum XOR gate as the
fundamental primitive operation of quantum logic is emphasized.
Recent developments in the use of quantum entanglement to built
error-robust quantum states, and the simplest protocol for
quantum error correction, are discussed.
\end{abstract}
\pacs{1996 PACS: 03.65.Bz, 07.05.Bx, 89.80.+h, 02.70.Rw}
% 02.20.Sv  Lie algebras of Lie groups
% 02.70.Rw  Other computational methods
% 03.65.Bz  (old) Foundations, theory of measurement, miscellaneous
% 03.65.Bz  (updated) Foundations, theory of measurement, quantum cryptography,
%           quantum computation, Aharonov-Bohn effect, Bell inequalities,
%           Berry's phase
% 03.65.Ca  Quantum mechanics: formalism
% 07.05.Bx  Computer systems: hardware, operating systems, computer
%           languages, and utilities
% 76.70.Fz Double nuclear magnetic resonance (DNMR), dynamical nuclear
%          polarization
% 89.80.+h Computer science and technology

\pagebreak

\section{Quantum gates and quantum spins}

I will give here a very brief summary here of the current activities in
the field of quantum computation, particularly some brand-new developments
on the preservation of quantum coherence in a noisy environment which
may be of interest to those working on the quantum physics of spin
systems.  It is an awkward story to tell at this point, since much of
it is in the process of creation as I write, and I cannot pretend that
we have a coherent picture of the whole at the moment.  But even the
incomplete story is, I think, of some considerable interest.

I will begin with some very basic points about quantum information, which
I suspect that many of you have seen before.  (There are a number of excellent
reviews available, see \cite{PT,mesci,EJ}.)
Suppose I imagine that I use
the states of one single, individual 
$s=1/2$ spin to represent the state of a bit;
why not?  (Of course, there are plenty of practical reasons why not, but
I will proceed anyway.)  I have to choose a data representation, but there
is a very obvious one: the spin-up state $|\uparrow\rangle$ may represent
the logical FALSE or zero state $|0\rangle$, and the spin-down state 
$|\downarrow\rangle$ may
represent the TRUE or one state $|1\rangle$.  Now, the state of a whole
register may be represented by the multi-spin state, say,
\begin{equation}
\Psi=|001010111...\rangle.
\end{equation}
This should already be a big pill to swallow for those of us in magnetic
physics; how do you put a system with many spins into {\em that specific}
wavefunction, and how do you make it stay there?  It just isn't done.
Well, it {\em ordinarily} isn't done; but the remarkable thing is that 
there are some areas of physics, like atomic physics experiments on ions in
vacuum traps\cite{ions}, 
where such multi-spin wavefunctions can be manufactured
and stabilized.  There is nothing in principle to forbid it in solid-state
or magnetic physics either, but it is rather exotic.  

Let me give an idea of how such multibit states $\Psi$ might be
produced in a solid state setting.  There will be a lot of
``suppose''s in this description, but if you don't like my supposes,
you may think up others for yourself.  There is one particular
multibit state which is easy to understand, and is in many contexts
easy to manufacture: the state of all zeros, $|000...\rangle$.  We may
get this by first putting the spins in a strong enough magnetic field,
and then letting them cool down.  In fact the all-zero state is a good
starting state for all quantum computation.  How, then, does one get
all the other bit states?  In the atomic physics realizations, and
perhaps someday in the solid state realizations too, this can be done
by manipulating the spins spectroscopically\cite{mesci}.  Suppose that
we can illuminate the spins with radiation which is tuned to the
resonant frequency connecting the lower, spin-down state with the
upper spin up state.  Further, suppose (and this is the hard
``suppose'') this can be done in a perfectly selective fashion, that
is, where {\em only one particular} spin is in resonance.  If you
complain that this is impossible, I refer you to the atomic
physicists, who are busy doing it.  One may imagine doing this by
focusing the radiation on a particular spin (if it is being done
optically), by applying some magnetic field gradient so that the
resonant frequency of some spin is uniquely shifted (this may perhaps
be doable with magnetic force microscopy), or by arranging that the
selected spin have a different chemical or crystallographic
environment (but this is hard to do in a general way for any spin in a
large collection).

If this selectivity has been achieved, then the idea is that the
radiation causes a controlled time-evolution of the spin's state;
in particular, it induces a Rabi oscillation between the lower and
upper states of the two-level system.  (Not coincidentally, we use
the language appropriate for coherent spectroscopy.)  If we allow the
system to undergo half of a Rabi oscillation period --- another jargon
for this would be that we have applied a $180^\circ$ 
tipping pulse to the state ---
then the state of that spin (suppose it is spin number 1) has been flipped:
\begin{equation}
\Psi=|100...\rangle
\end{equation}
Note that this process is perfectly reversible; if we subject the system
to another half-period of Rabi oscillation, it will return to the
original state.

So far, this discussion has not used the ``quantumness'' of the spin
states at all; the states discussed above have a one-to-one correspondence
to the states in an ordinary computer register.  If this were all that
there was to it, quantum computing would be singularly uninteresting, it
would merely be a massively difficult way of mimicking the states of
an ordinary computer.  What makes things interesting is that one can
concoct quantum states which have no analog on a classical machine.
Consider what happens if the resonant radiation illuminating spin number
1 is only left on for {\em one-quarter} of a Rabi oscillation (corresponding
to a $90^\circ$ tipping pulse).  Then the state of the system evolves into
\begin{equation}
\Psi=\frac{1}{\sqrt{2}}|000...\rangle+
     \frac{1}{\sqrt{2}}|100...\rangle
\end{equation}
a coherent superposition of two bit states.  Obviously, the memory of an
ordinary computer cannot be placed into a superposition of values; we often
remind ourselves of this important difference by using the term {\em qubit} 
rather than bit to refer to this object.   Of
course, the computer may be programmed to {\em represent} such superpositions,
but this is very different from the computer memory {\em being} in this
superposition.  Also, this ``software'' approach is massively inefficient
--- it is well known that it takes computer resources proportional to 
$\exp(n)$ to represent an $n$-bit quantum state. 
It is the possibility of
making such superposition states that lends quantum computers their
great potential power, permitting them to do parallel computations
that make certain problems, like prime factorization, solvable in
totally new ways\cite{Shor}.  I will say no more about this here, 
but refer the
reader to various excellent writeups of these results.

But I do want to say something about another obvious question that
this brings up: how do you do logic, how do you implement logic gates,
which act on the states of these spins?  It turns out that besides the
one-bit rotations discussed above, only one other gate, a two-bit
logic gate which is variously known as the controlled-NOT gate or
the quantum XOR gate, is the only additional ingredient which is
needed to do logic on quantum states\cite{G9}.  
The controlled-NOT gate has
a very classical description: it flips the state of one bit (the
{\em target} bit) contingent upon the state of a second ({\em
control} bit).  That is, it performs the following transformation
on the four possible initial states of two bits:
\begin{equation}
\begin{array}{lll}
00&\rightarrow&00\\
01&\rightarrow&01\\
10&\rightarrow&11\\
11&\rightarrow&10\end{array}
\end{equation}
Here the first bit is the control, the second is the target.  Fig.
1 shows the conventional gate symbol for this operation.  

While the controlled-NOT's specification is classical, its implementation
cannot be, for it must act on quantum bits, and it must therefore preserve
any quantum superpositions with which it is presented, so that, for
example, the controlled NOT must successfully perform the following
transformation of a superposed input:
\begin{equation}
a|00\rangle+b|11\rangle\rightarrow a|00\rangle+b|10\rangle
\end{equation}
That is, each member of the quantum superposition must be transformed
according to the classical rule for the gate.

It will not surprise the reader that implementing this gate will take
a few more ``suppose''s.  The main additional ingredient which is required
is an interaction between the spins, that is, a non-zero matrix element
of the Hamiltonian coupling them.  One may imagine a system which is
engineered so that this interaction is turned on temporarily when the
gate needs to act (this is the basis of some other proposals I have
made using magnetic force microscopy\cite{mesci,divi}; 
alternatively, fixed interactions,
provided they have special, simple forms, will also suffice.  For the
multi-spin systems, an Ising-like interaction, one in which the
coupling interaction only involves the $\sigma_z$ Pauli matrix:
\begin{equation}
H_{int}=J\sigma_z^i\sigma_z^j
\end{equation}
is satisfactory.  One important reason why this special form of the
interaction Hamiltonian is desirable is that all the classical bit
states (e.g., $|00101011...\rangle$ are eigenstates of this interaction,
and so there is no undesirable time evolution of the quantum states.
However, a time evolution corresponding to the controlled-NOT can
be induced in this system, if again we suppose that we can illuminate
the system with radiation which is focussed on just the target spin
for the controlled-NOT.
Because of the interaction, the transition frequencies 
depend on the joint state of the two particles.  In particular, the
$|00\rangle\rightarrow|01\rangle$ transition is at energy $E_Z+J$,
while the transition energy for $|10\rangle\rightarrow|11\rangle$
has energy $E_Z-J$, where $E_Z$ is the Zeeman splitting caused by
a uniform magnetic field.  Because of the shift caused by $J$, 
the radiation can
be tuned so as to be in resonance with one transition and not the other.
Except for being specific to one unique spin, the use of such a splitting
which is produced by another, coupled quantum system
is the common basis of all of double-resonance spectroscopy.  It also
suffices, if one-half of a Rabi oscillation is induced, to perform 
essentially the desired controlled-NOT operation.  (The phases of the
induced transition require a careful discussion, which has been given
elsewhere\cite{Cleve}.  
This discussion does not change the essential features of
the argument given here.)

\section{Noisy quantum systems and error correction}

I have briefly indicated above the kinds of quantum-state manipulations
that are necessary in order to perform quantum computations.  I now wish to 
touch upon some of the highlights of an area of intense interest in quantum
computation theory now, an area which is absorbing my attention and the
attention of many other workers in the field.  This is the subject
of ``quantum error correction'', which involves asking the question,
how imprecisely can the above operations be performed and still be useful
for quantum computation?

Imprecision in the manipulation of quantum systems comes in two apparently
very different forms, which however can really be treated in the same way in
quantum error correction.  They are ``decoherence'' and ``mis-rotation''.
Decoherence refers to what happens to the quantum system in the presence
of an environment; I would like to make a few comments about the mathematics
which we use to describe this situation.  When a quantum system is {\em not}
interacting with its environment, its wavefunction 
$\Psi$ evolves in time according
to some unitary time evolution operator $U$:
\begin{equation}
\Psi_{final}=U\Psi_{initial}.
\end{equation}
A completely equivalent description is that the system's {\em density matrix}
$\rho=|\Psi\rangle\langle\Psi|$ evolves as
\begin{equation}
\rho_{final}=U\rho_{initial}U^\dagger.\label{den1}
\end{equation}
It is unnecessary to use the density-matrix description for an isolated
quantum system, it {\em is} necessary for an open quantum system which is
dephasing by interacting with its environment.  In this case there is a
description of the time evolution which is analogous to Eq. (\ref{den1}),
which can in general be written
\begin{equation}
\rho_{final}=\sum_iA_i\rho_{initial}A_i^\dagger.\label{den2}
\end{equation}
The set of matrices $\{A_i\}$ together make up what is known as a {\em
superoperator}.  The superoperator description of noisy quantum systems
is not widely used, but it should be; it is a concise and precise 
description of the quantum environment, and its parameters may be
determined by measurement.  (I highly recommend the paper of 
Schumacher\cite{Shoe} as a compendium of properties of superoperators.)

We will see various examples of superoperator descriptions of the noisy
environment below, but I would first like to mention a very important
example of one, which is known as the ``depolarizing channel''\cite{mongo}.  
To call
an environment a ``channel'' borrows some notation from the theory of
(quantum) communications, where a qubit is deemed to be subjected to noise
during the time that it is resident in the ``channel'', in motion from
place to place; this is only a matter of nomenclature.  The superoperator
for the depolarizing channel is specified by the four matrices:
\begin{equation}
\begin{array}{rr}A_0=\sqrt{f}\left(\begin{array}{rr}1&0\\0&1\end{array}\right)&
A_1=\sqrt{\frac{1-f}{3}}\left(\begin{array}{rr}1&0\\0&-1\end{array}\right)\\
\ &\ \\
A_2=\sqrt{\frac{1-f}{3}}\left(\begin{array}{rr}0&1\\1&0\end{array}\right)&
A_3=\sqrt{\frac{1-f}{3}}\left(\begin{array}{rr}0&1\\-1&0\end{array}\right)
\end{array}
\end{equation}
The effect of this channel (or superoperator) is to leave the quantum
state alone with probability $f$, and to rotate it about one of three
orthogonal axes by $90^\circ$ (I am using a ``Bloch sphere'' notation
here) with probability $(1-f)/3$.  Another equivalent description is
that it involves leaving the quantum state alone with probability 
$(4f-1)/3$, and, with probability $(2-4f)/3$, replacing it with a 
completely random quantum state.  ($(2-4f)/3$ is referred to as the
depolarization fraction in this case.)

Now, after all this, let me return to the question of imprecision of
quantum gate operations.  Suppose that because of the nature of your
experimental setup, when you are trying to make a gate corresponding to
the unitary transformation $U(\{\theta_0\})$ with some specified 
rotation
angles $\{\theta_0\}$ you will actually produce one of a range of 
unitary transformations drawn from a probability distribution
$p(\{\theta_0\})$.  Then the new density matrix after operation of
this partially known gate is
\begin{equation}
\rho_{final}=\int d\{\theta\} p(\{\theta\})
U(\{\theta\})\rho_{initial}U^\dagger(\{\theta\})\label{den3}
\end{equation}
There is clearly a correspondence between the matrices 
$\sqrt{p(\{\theta\})}U(\{\theta\})$ and the matrices $A_i$ 
of the superoperator
above.  The correspondence may be made even more exact with a 
Gram-Schmitt transformation which converts the
continuous set of matrices $\sqrt{p(\{\theta\})}U(\{\theta\})$ into
a discrete set --- just four of them, in fact, when the dimension
of $\rho$ is 2, just as in the depolarizing-channel example
above.  The conclusion of this is that uncertainty in the
unitary transformation performed by a quantum gate is really
indistinguishable from the situation in which first the gate is performed
perfectly, and then the qubits interact with some noisy environment.
This effective environment can have some strange properties;
for instance, for the case of an uncertain two-bit gate, the
effective noise may correspond to some correlated interaction
of the two qubits with the effective environment, rather than
the common case of a real environment in which the two qubits
interact independently with their environments (we will see in
a moment a superoperator description of such a situation).

So, decoherence and misrotations are types of errors in quantum
computation which can be placed on exactly the same footing. 
Now, we come to the question of how one is to correct for such
errors.  In less than a year preceding this writing, the theory of
error-correction of quantum systems has gone from nearly nothing
to a fantastically developed and mathematically sophisticated
edifice\cite{biggie}.  This monumental theoretical development also has,
in my opinion, the potential to influence experimental quantum
physics (not just quantum computation) in many profound ways.

Having said this, I must humbly admit that I will give here only a
few small morsels of the feast which appears to await us.  I think
that theorists are still struggling to explain all that they
have discovered about quantum error correction, and an attempt
at a full explanation of current developments would come across
as abstruse and obscure, at least if I were to write it.  But
in offering up only morsels I have the option of explaining
just the most elementary parts of the ideas, and also the ones
which are likely to most readily testable within the current
experimental art.  

In this I hope to bring home a rather paradoxical idea about entangled
states.  Entangled states, famous in quantum physics in the
Einstein-Podolsky-Rosen paradox\cite{EPR}, Bell's
inequalities\cite{Bell}, and so forth, are quantum states of two (or,
especially for the purposes of quantum error correction, more than
two) subsystems, in which the wavefunction of the pair is not a
product of the wavefunctions of the individual subsystems.  For two
spin-1/2 systems, for example, a legal quantum state is
\begin{equation}
\Phi=|00\rangle+|11\rangle.
\end{equation}
(Recall that we denote spin down by 0, spin up by 1.)
This state and other related ones are illustrative
of the weird non-locality of quantum correlations; when the
two spins are spatially separated, we say that the {\em common}
direction of the two separate spins is perfectly correlated,
even while each separately has a completely indeterminate direction.
(We will not be so much interested, in the examples from quantum
error correction theory, of the weirdness attendant on the remote
physical separation of these components.  We will be content, in
fact, for all the components to be in close physical proximity.
The mathematical weirdness persists.)  The EPR-type state above
is also the smallest example of a ``macroscopic'' superposition
state; the really macroscopic version of this is 
the famous ``Schrodinger cat'' state\cite{SC}, 
$|$alive$\rangle$+$|$dead$\rangle$. 

Both of the states illustrated above are considered to be very fragile
quantum states; a small amount of interaction with the environment
changes the state into a classical mixture in which all of the
spooky quantum correlations have been lost.  But, the paradoxical
message from quantum error correction theory is that while entangled
states {\em do} fall apart when subjected to noise from the environment,
they can be designed to fall apart in such a way that the quantum
correlations are not {\em irretrievably} lost, and so that the
original quantum state is in fact {\em recoverable} (via a quantum
computation, of course).

With this run-up, I will finally give my one simple worked-through
example of quantum error correction which illustrates all of the 
assertions I have made below.  Suppose the object is simply to
store (or perhaps transmit) a quantum bit, which may be in some
arbitrary state
\begin{equation}
\Psi=a|0\rangle+b|1\rangle\label{super}
\end{equation}
i.e., with arbitrary complex coefficients $a$ and $b$ (satisfying the
normalization condition).  Now, I imagine that this qubit is subjected
to a (toy) environment characterized by the two matrices
\begin{equation}
A_1=\sqrt{f}\left(\begin{array}{rr}1&0\\0&1\end{array}\right)
\end{equation}
\begin{equation}
A_2=\sqrt{1-f}\left(\begin{array}{rr}0&1\\1&0\end{array}\right)
\end{equation}
$f$ is a characterization of the fidelity of this ``channel'', with
$f=1$ corresponding to the noiseless case.  It is a toy channel
because it does not correspond the typical real noisy channel;
in words one can say that with probability $f$ it leaves the qubit
alone, and with probability $1-f$ is flips 0 to 1 and {\em vice 
versa}.  As one can see, it is a stripped-down version of the
depolarizing channel above (for some comments about the error
correction for that case, see the very end); also, it bears some
resemblance to a ``decay'' channel, which is a very real
model for some applications in quantum optics\cite{ChuangLaf}, 
where the superoperator is specified by
\begin{equation}
A_1=\left(\begin{array}{rr}1&0\\0&\sqrt{f}\end{array}\right)
\end{equation}
\begin{equation}
A_2=\left(\begin{array}{rr}0&\sqrt{1-f}\\0&0\end{array}\right)
\end{equation}
where typically $f$ is taken to be an exponential function of
time: $f(t)=e^{-t/\tau}$, with some characteristic decay time
$\tau$.  Despite their resemblances, though, the decay channel
takes rather more work to error-correct than my toy channel,
and I will be sticking with that here.  

So, enough of the environment.  What does it do to my poor qubit?
Of course, that depends on what $a$ and $b$ are; in general, though,
I will characterize the quality of the degraded qubit by a 
fidelity, an overlap between the original state and the state
after interaction with environment.  If the density matrix
specifying this final state is $\rho$, then the fidelity $F$
is simply
\begin{equation}
F=\langle\Psi|\rho|\Psi\rangle.
\end{equation}
$F$ is obviously a function of $a$ and $b$, as well as the
channel parameter $f(t)$ (or just $t$).  The right thing to do
is to minimize $F$ over $a$ and $b$, or take averages over their
expected distribution, but I will just pick particular $a$'s
and $b$'s and discuss the behavior of $F$ with $t$.

It is trivial to see that if $a=1$ in Eq.~\ref{super}, then
\begin{equation}
F(t)=e^{-t/\tau}
\end{equation}
i.e., the fidelity of the state falls exponentially.  Now we
will see how the tools of quantum error correction can be used to
improve upon this.  The first idea is that we replace the bare
qubit with a coded qubit.  In this case a good way to code the
qubit is to represent the $|0\rangle$ state by the three-qubit
state $|000\rangle$, and the $|1\rangle$ by the state $|111\rangle$;
thus, the qubit is encoded in the three-qubit entangled state
\begin{equation}
\Psi=a|000\rangle+b|111\rangle
\end{equation}
This state is a close relative of the ``Greenberger-Horne-Zeilinger''
state\cite{GHZ}, which has long been discussed in scenarios of
multi-particle entanglement and quantum non-locality.  Now we will see
how this state can be used to more-robustly (i.e., with higher
fidelity) preserve the qubit against the ravages of the noisy channel.

First, we imagine that each of the three qubits in the state above
is subjected to an independent copy of the toy channel introduced
above.  This means that the operators which make up the complete
superoperator will have a direct-product form:
\begin{equation}
A_{ijk}=A_i^{(1)}\otimes A_j^{(2)}\otimes A_k^{(3)}.
\end{equation}
If we compute the fidelity after this superoperator has been applied,
just
using the same formula as above, we find that it is actually {\em
less} than for the uncoded qubit, in fact, again for the case 
$a=1$ in Eq.~(\ref{super}):
\begin{equation}
F(t)=e^{-3t/\tau}
\end{equation}
That is, the fidelity falls three times faster than for the unencoded
state; so, entangled states {\em are} more delicate than unentangled
ones.  

We seem to be not at all closer to our goal of making the qubit last
longer.  But the key is that, while the noisy entangled state has
lower fidelity, it is more {\em recoverable}.  To see what this means,
suppose that the toy channel had only acted on one of the qubits, and
that the qubit had been flipped.  In this case, the state of the system
may have been turned into
\begin{equation}
\Psi=a|100\rangle+b|011\rangle.
\end{equation}
(The full density matrix of the state after it has interacted with
the channel is a statistical mixture of such flipped states with
appropriate probability weights.)  Now, this state has {\em zero}
fidelity, because it is orthogonal to the original coded qubit
state.  However, unlike the uncoded qubit, the coded state can
be processed, by an {\em error correction} step, to recover the
original qubit exactly.  (Details of the example which I show now
has been presented by A. M. Steane in Ref.~\cite{biggie}.)

This error correction proceeds in the following fashion: We perform a
sequence of incomplete von Neumann measurements on the state.  The
first one asks the question, ``are qubits 1 and 2 in the same state,
or not?''  To ``ask'' this question of the quantum state, we perform a
quantum computation on it, using (not surprisingly) the quantum XOR or
controlled NOT operation which I emphasized in the first part as the
fundamental gate of quantum computation.  Here's how it works
(referring to Fig. 2):
introduce a fourth, ancillary bit, initially setting it to
$|0\rangle$.  Then do two XORs, the first with qubit 1 as the source
bit, the second with qubit 2 as the source; in both cases, the target
is the ancilla qubit 4.  If qubits 1 and 2 are in the state 00, then
bit 4 is untouched; if they are in the state 11, then qubit 4 is
flipped twice, and so is again in its original state $|0\rangle$.  On
the other hand, if the source state is 01 or 10, the ancilla bit gets
flipped just once, and ends up in the state $|1\rangle$.  Then, a
measurement of qubit 4 is exactly the incomplete measurement we have
specified above.  It is crucial to recognize that this measurement
obtains information about the entangled state without disturbing it.
This is only possible because of the special form of the entangled
state; the two parts of the state $a|100\rangle$ and $b|011\rangle$
give the same measurement outcome, so their quantum superposition is
undisturbed.

Just one other measurement of this sort, on qubits 1 and 3, suffices
to determine exactly which of the three qubits was disturbed by the
channel.  Referring to Fig. 2, it is easy to see that if
$M_1=M_2=0$, then none of the three qubits is disturbed, if 
$M_1=1$ and $M_2=0$, then it must have been qubit 1 that had been
flipped, if $M_1=0$ and $M_2=1$ then the disturbed qubit must have
been number 3, and if $M_1=M_2=1$, the number 2 is the disturbed one.
Since the disturbance caused by the channel is completely known, and the
form of the entangled state is undisturbed by these measurements,
A final quantum gate performing a flip or NOT on the affected
qubit serves to restore the state exactly to its original form.

Of course, this error correction procedure, just like any classical
one, will be fooled if too many qubits have been flipped.  The
procedure above will be fooled if a two-bit error is caused by
interaction with the environment.  But at early times, when
the probability of a single qubit-flip $(1-f)$ is small, the
probability of two qubit flips is much smaller, of order $(1-f)^2$.
Therefore, if the error correction is performed after a sufficiently
brief interaction with the environment, the error-corrected state
will have higher fidelity than the bare qubit state.  Fig. 3 shows
these fidelities for the example above; the coded state has higher
fidelity up until a time corresponding to $f=1/2$.

The crucial feature of these curves, though, is that the early-time
fidelity is very high, going like $1-c(1-f)^2$.  This means that if
error correction is done often enough (more frequently than the 
natural decay time), the fidelity may be kept high for much longer
than would be possible for the uncoded qubit in contact with its
environment.  This is the essence of the error-robustness of entangled
states.

I have finished the tale I wanted to tell, but it is really just the
tip of the iceberg on the story of quantum error correction, which
I and a number of other workers in the field have been intensely
studying.  
First, we know of encodings that will protect states, in
the same way just outlined, from more general types of interaction with
its environment.  In particular, we\cite{mongo} 
and others found a code that would
protect against interaction with the depolarizing channel above; five-bit
entangled states are required for this.  This code also works for the
quantum-optics decay channel, although it is not known whether there
might be more efficient codes for dealing with this particular case.
There are also large families of codes which are capable of detecting
and correcting more than one-bit errors\cite{biggie}.  
This means that the short-time
corrected fidelity is better than the example given above, going like
$F(t)=1-c(1-f)^{t+1}$, where $t$ is the number of errors corrected.
Larger-size entangled states, and more extensive quantum computation
for error correction, are required for this.  Still, there is a
recent idea of Shor\cite{Shorft}, 
being actively investigated by a number of workers,
that large-block codes, correcting many errors, may be optimal for 
performing long quantum computations reliably.  There are, I believe,
several years of theoretical work still ahead of us to find the
ultimate capability of quantum computation, particularly in the
presence of noise.

\pagebreak
\begin{figure}
\vspace{2cm}
\epsfbox{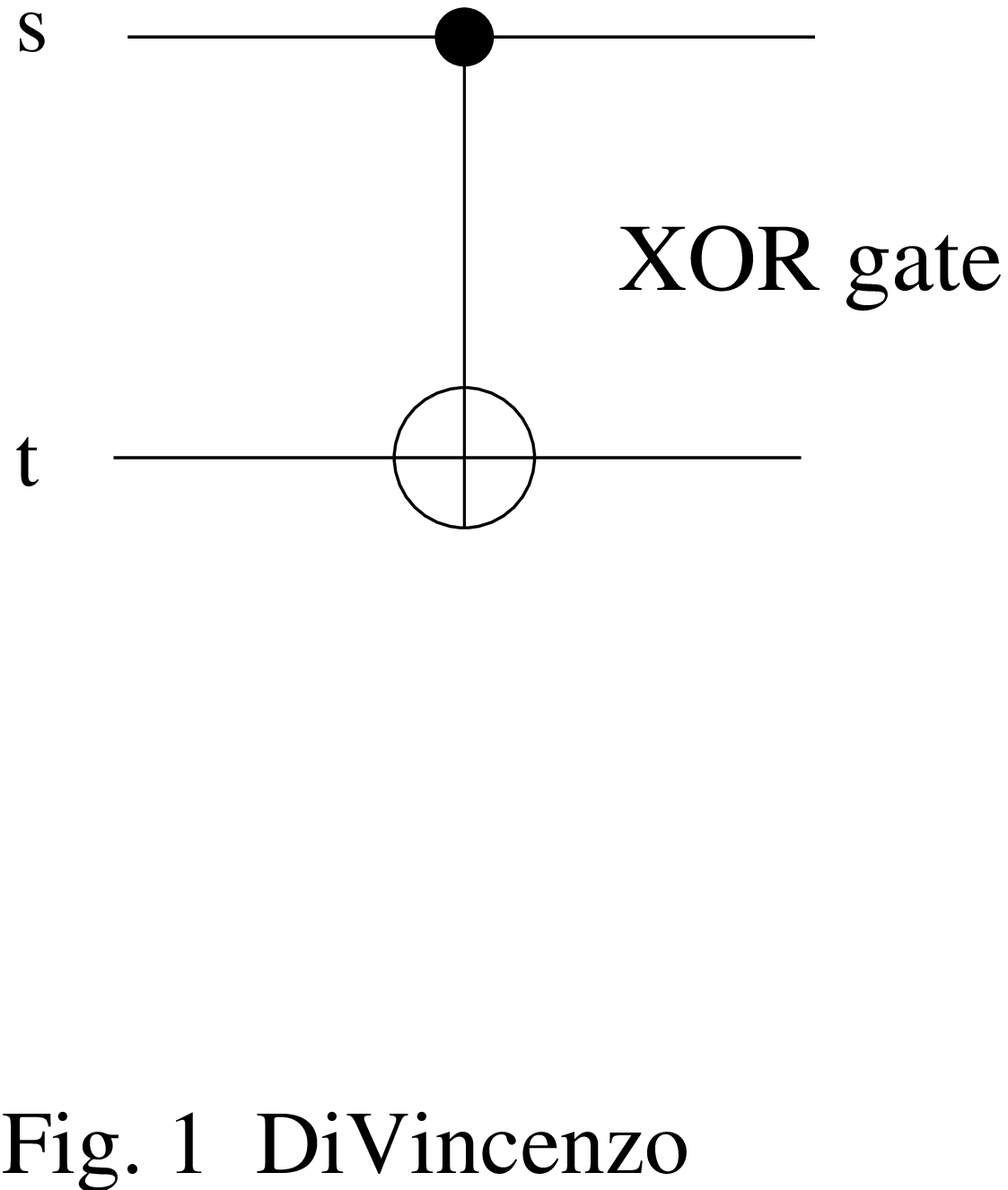}
\vspace{2cm}
\caption{The gate symbol for the controlled-NOT or XOR gate, the
fundamental two-bit gate of quantum logic.  The upper line denotes
the source qubit, and the lower line the target qubit.}
\end{figure}

\pagebreak
\begin{figure}
\vspace{2cm}
\epsfbox{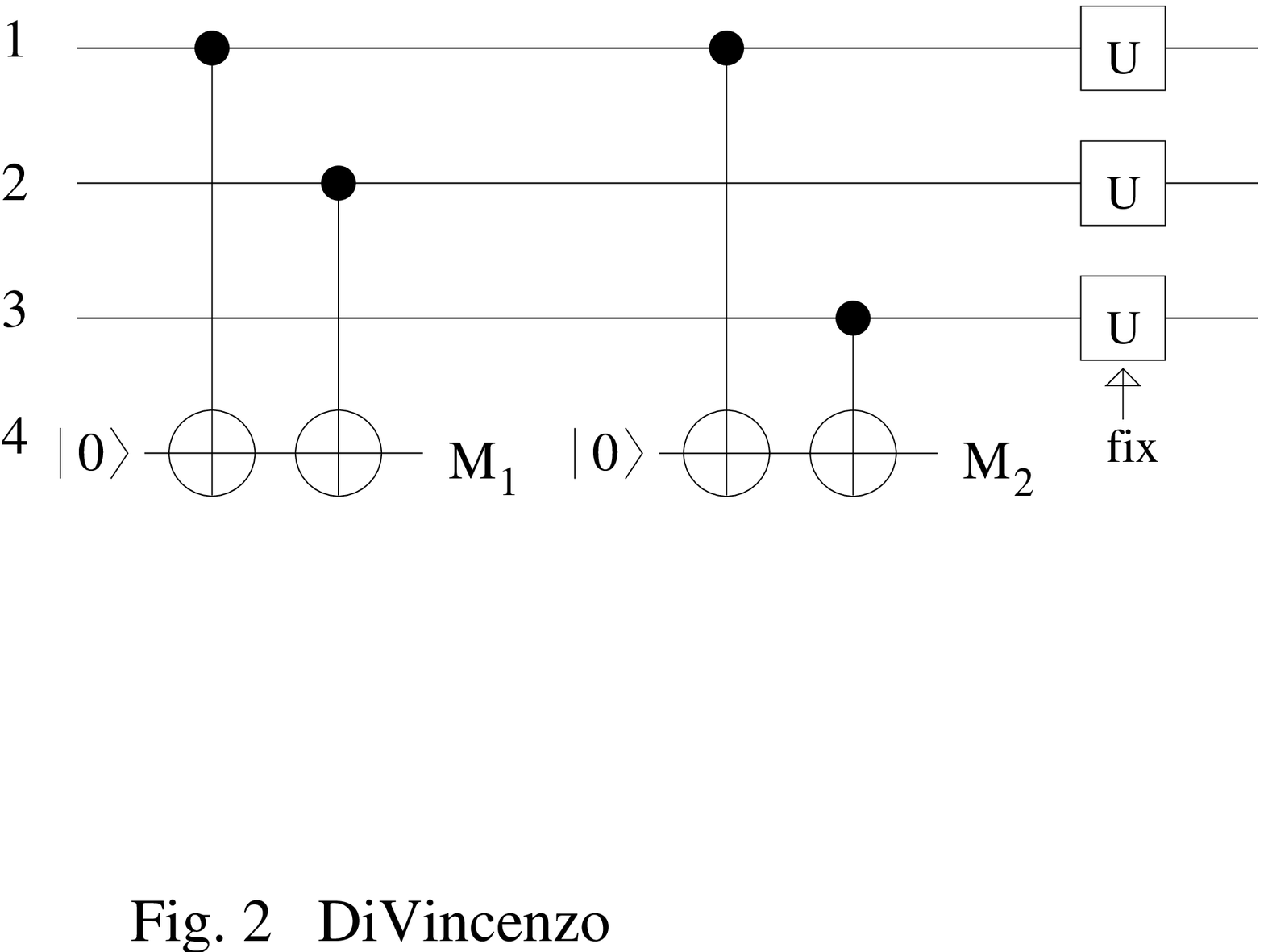}
\vspace{2cm}
\caption{Quantum network, using XOR gates, one-bit measurements, and
a final one-bit rotation, for performing error correction on the entangled
code state.}
\end{figure}

\pagebreak
\epsfbox{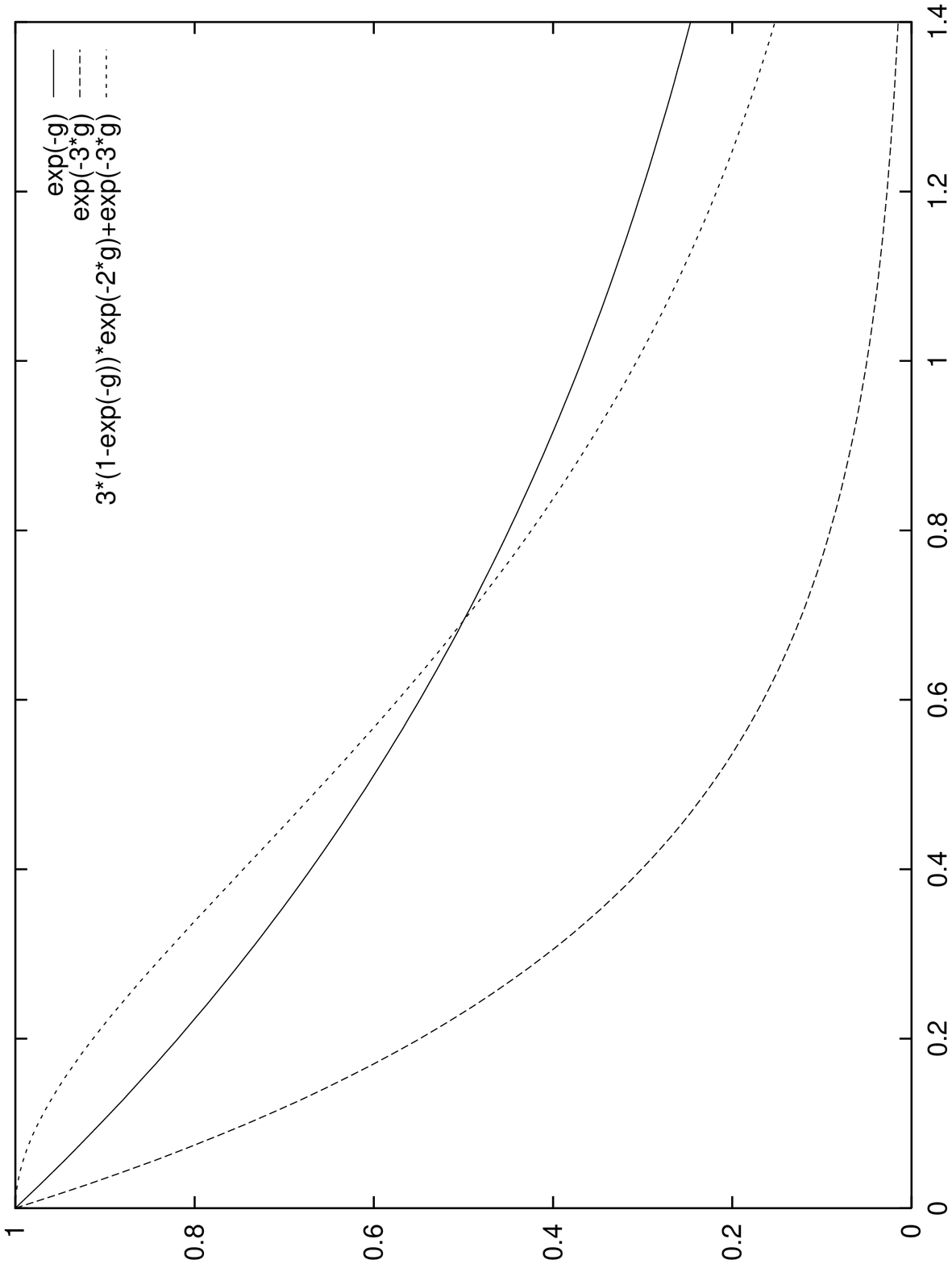}

\pagebreak
\begin{figure}
\caption{Fidelity {\em vs.} time with and without quantum error correction,
as described in the text.  Solid line: $f=\exp(-t/\tau)$, fidelity of the
unencoded qubit.  Dashed line: $f=\exp(-3t/\tau)$, fidelity of the
triple-coded qubit without error correction.  Dotted line: $f=3(1-\exp(-t/
\tau))\exp(-2t/\tau)+\exp(-3t/\tau)$, the fidelity of the coded state
after it has been error-corrected at time $t$.}
\end{figure}

\end{document}